\documentclass[
 preprint,
 amsmath,amssymb,
 prl,
 superscriptaddress,
]{revtex4-1}
\usepackage{graphicx}
\usepackage{dcolumn}
\usepackage{bm}

\begin{document}
\preprint{APS/123-QED}
\thanks{A footnote to the article title}%
\title{Enhancement of UV Second-Harmonic Radiation at Nonlinear Interfaces with Discontinuous Second-order Susceptibilities}
\author{Xiaohui Zhao}
\affiliation{Department of Physics and Astronomy, Shanghai Jiao Tong University, 800 Dongchuan Road, Shanghai 200240, China}
\author{Yuanlin Zheng}
\email{ylzheng@sjtu.edu.cn}
\affiliation{Department of Physics and Astronomy, Shanghai Jiao Tong University, 800 Dongchuan Road, Shanghai 200240, China}
\author{Huaijin Ren}
\affiliation{Institute of Applied Electronics, China Academy of Engineering Physics, Mianyang, Sichuan 621900, China}
\author{Ning An}
\affiliation{Shanghai Institute of Laser Plasma, China Academy of Engineering Physics, Shanghai, 201800 China}
\author{Xuewei Deng}
\affiliation{Laser Fusion Research Center, China Academy of Engineering Physics, Mianyang, Sichuan 621900, China}
\author{Xianfeng Chen}
\email{xfchen@sjtu.edu.cn}
\affiliation{Department of Physics and Astronomy, Shanghai Jiao Tong University, 800 Dongchuan Road, Shanghai 200240, China}
\date{\today}
\begin{abstract}
We investigate the generation of ultraviolet (UV) second-harmonic radiation on the boundary of a UV transparent crystal, which is derived from the automatic partial phase matching of the incident wave and the total internal reflection. By adhering to another UV non-transparency crystal with larger second-order nonlinear coefficient $\chi^{(2)}$, an nonlinear interface with large disparity in $\chi^{(2)}$ is formed and the enhancement of UV second-harmonic radiation is observed experimentally. The intensity of enhanced second harmonic wave generated at the nonlinear interface was up to 11.6 times at the crystal boundary. As a tunable phase-matching method, it may suggest potential applications in the UV, even vacuum-UV, spectral region.

\end{abstract}
\maketitle

Coherent light sources at ultraviolet (UV) and vacuum ultraviolet (VUV) delivered by all solid-state laser systems are highly demanded due to numerous practical applications in lithography \cite{suganuma2002157}, photoelectron spectroscopy \cite{togashi2003generation,kiss2005photoemission,kiss2008versatile}, micro-machining and semiconductor processing \cite{okazaki2012octet}, high-density optical storage \cite{pudavar1999high} and so on. Seeking for suitable transparent nonlinear media for UV and VUV harmonic wave generation has attracted great interest over past decades. Currently, the most potential crystals to produce VUV coherent light by second-order nonlinear processes are KBe$_2$BO$_3$F$_2$ (KBBF) \cite{chen1995new,wu1996linear,chen2004recent,kanai2004generation,nakazato2016phase} and BaMgF$_4$ (BMF) \cite{shimamura2005advantageous,kannan2008ferroelectric,shimamura2011growth} with cutoff wavelengths in the UV region of 153 nm and 126 nm, respectively.

With relatively large birefringence, the limit for birefringent phase matching (BPM) of second-harmonic generation (SHG) in KBBF is 160 nm, and the shortest wavelength has achieved by sum-frequency generation is 153.4 nm \cite{nomura2011coherent}. However, the layered structure is a big limitation of KBBF, which restricts the crystal size, cutting angles and so on. Avoiding of the walk-off effect and other shortcomings of BPM, qusai-phase matching (QPM) pattern for SHG \cite{vyllora2009birefringent,mateos2014bamgf4} is permitted by the ferroelectricity of BMF. But the fabrication of periodically revered domain structures in bulk media with a period less than 3 $\mu$m leastwise \cite{Buchter:01}, which is the requirement of SHG in the VUV wavelength region, is a big challenge. So far, QPM has not been demonstrated in the UV with BMF. In addition, the nonlinear coefficients of these two VUV-transparent crystals ($d_{11}$=0.49 pm/V for KBBF and $d_{32}$=0.039 pm/V for BMF) are significantly smaller than that of the nonlinear crystals commonly employed in visible to infrared region, for example LiNbO$_3$ (LN) ($d_{33}^{\mathrm{LN}}$=34.45 pm/V).

To overcome these difficulties, other efficient frequency conversion processes taking advantage of tunable and flexible phase matching methods can find a way out. Derived from complete phase matching of the incident light and the scattering light, conical SHG could be observed in bulk media \cite{ren2013surface,Li:15}. The triangle phase-matching relationship could be realized as $\vec{k_1}+\vec{k_1^\prime}=\vec{k_2}$, where $\vec{k_1}$ and $\vec{k_2}$ are the wave vectors of fundamental wave (FW) and second harmonic (SH) in the medium, respectively. The additional FW vector $\vec{k_1^\prime}$ is provided by the scattering light. Taking advantage of total internal reflection on the crystal boundary, $\vec{k_1^\prime}$ could be replaced by reflected FW, which can greatly improve the SHG efficiency. However, these triangle phase-matching methods also have a limit for wavelength, due to the dispersion relation of crystal.

\begin{figure}[htb]
\centerline{
\includegraphics[width=6 cm]{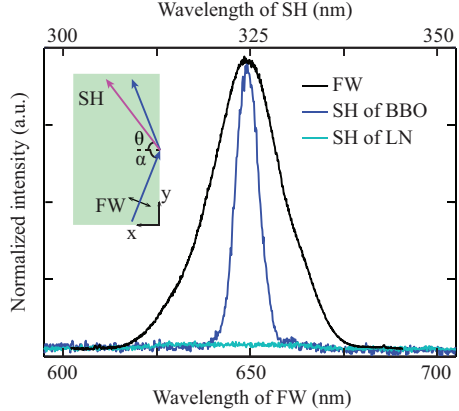}}
\caption{Spectrum of the FW and harmonic wave generated at the boundary of BBO and LN. The center wavelength of FW and SH is 650 nm and 325 nm, respectively. The inset is experimental schematic of the generation of UV second-harmonic wave generated on the boundary of the BBO (or LN) crystal.}
\end{figure}

In this letter, we report on an automatic non-collinear phase matching method, which exploits the internal total reflection but has no limit for wavelength. The SH radiation was enhanced at the nonlinear interface with large disparity in $\chi^{(2)}$, which was formed by adhering to another UV non-transparency crystal with higher $\chi^{(2)}$. And the added crystal doesn't change the properties of SH radiation, for example the emission angle.

At first, a x-cut $\beta-BaB_2O_4$ (BBO) sample (absorption edge is 189 nm, $d_{11}^{\mathrm{BBO}}=2.26$ pm/V) of the size of 5 mm(x)$\times$10 mm(y)$\times$10 mm(z) was employed as the transparent nonlinear medium to generated the UV SH wave. The light source we used was an optical parametric amplifier (TOPAS, Coherent Inc.), producing 80 femtosecond pulses (1000 Hz rep. rate) at the variable wavelengths from 280 nm to 2600 nm. The FW was adjusted to be ordinary polarized with the wavelength of 650 nm and loosely focused by a 100-mm focal length lens to a beam waist of ~50 $\mu$m. The spectrum of the FW was shown in Fig. 1.

The inset of Fig. 1 shows the experimental schematic: the beam was obliquely incident into the BBO crystal and reflected on the boundary with an angle $\alpha$ and the output light was then collected by a fiber optic spectrometer. As the FW reflection, the stimulated nonlinear polarization wave propagates along the BBO boundary with a wave vector $k_{np}=2k_1\mathrm{sin}\alpha$. When the wave vector of nonlinear polarization smaller than that of harmonic, SH wave emits at a specific angle $\theta$ which is regarded as the nonlinear Cherenkov radiation satisfying the longitudinal phase-matching condition \cite{zhang2008nonlinear,ren2012nonlinear,ren2013enhanced}: $k_2\mathrm{sin}\theta=2k_1\mathrm{sin}\alpha$. Since this is a automatic and tunable phase-matching process, one can always find suitable incident angles for SHG without wavelength limitation. In experiment, we observed SH generated at the BBO boundary and the spectrum was shown in Fig. 1. As a comparison, a z-cut periodically poled 5 mol\% MgO:LiNbO$_3$ (LN) sample with the size of 10 mm(x)$\times$10 mm(y)$\times$5 mm(z) was employed to repeat the above experiment. Although the second-order nonlinear coefficient of LN sample is significantly higher than that of BBO, the SH was completely absorbed as the cutoff wavelength is 380 nm in LN.

To analyse the properties of SH generated at the boundary of BBO, a more general situation was considered: $y=0$ plane is a interface with nonlinear coefficient $\chi_1^{(2)}$ and another medium with $\chi_2^{(2)}$ [see Fig. 2(a)]. The coupled wave equation of this nonlinear process was solved using the Fourier transform \cite{sheng2012role,Zhao:16}, and the intensity of SH $I_2$ can be expressed as
\begin{multline}
I_2(k_x,y)=\Big(\frac{k_2}{2n_2^2}\Big)^2I_1^2y^2\mathrm{sinc}^2\Big[\frac{(\Delta k-k_x/2k_2)y}{2}\Big] \\
\Big|[\chi^{(2)}_1+\chi^{(2)}_2]\sqrt{\frac{\pi}{8}}ae^{-\frac{a^2k_x^2}{8}}+i[\chi^{(2)}_1-\chi^{(2)}_2]\frac{\sqrt{\pi}}{2}aD\big(\frac{ak_x}{2\sqrt{2}}\big)\Big|^2,
\end{multline}
where $k_x$ is the components of $k_2$ in $x$ directions, $n_2$ is the refractive index of the SH, $I_1$ denotes the intensity of the gauss-type FW with beam width $a$. The phase mis-matching is $\Delta k=k_2-k_{np}$, and $D(\frac{ak_x}{2\sqrt{2}})$ denotes the Dowson function.

According to Eq. (1), one can find that $I_2$ will have a maximum when the phase-matching condition $\Delta k=k_x/2k_2$ is satisfied.  The radiation angle of SH can be derived as $k_2\mathrm{sin}\theta=2k_1\mathrm{sin}\alpha$, which is exactly the longitudinal phase-matching condition. The absolute value $|\chi^{(2)}_1-\chi^{(2)}_2|$ directly affects the SH intensity, since the term containing $e^{-a^2k_x^2/8}$ decreases sharply with the increasing of the absolute value of $k_x$. When the medium is air, $\chi_2^{(2)}$ equals to 0, representing the condition of the above experiment that SH generated at the boundary of BBO. If the nonlinear coefficient of another medium is large enough that the relation $|\chi^{(2)}_1-\chi^{(2)}_2|>|\chi^{(2)}_1|$ is satisfied, the SH intensity is thus enhanced theoretically.

\begin{figure}[htb]
\centerline{
\includegraphics[width=12.0cm]{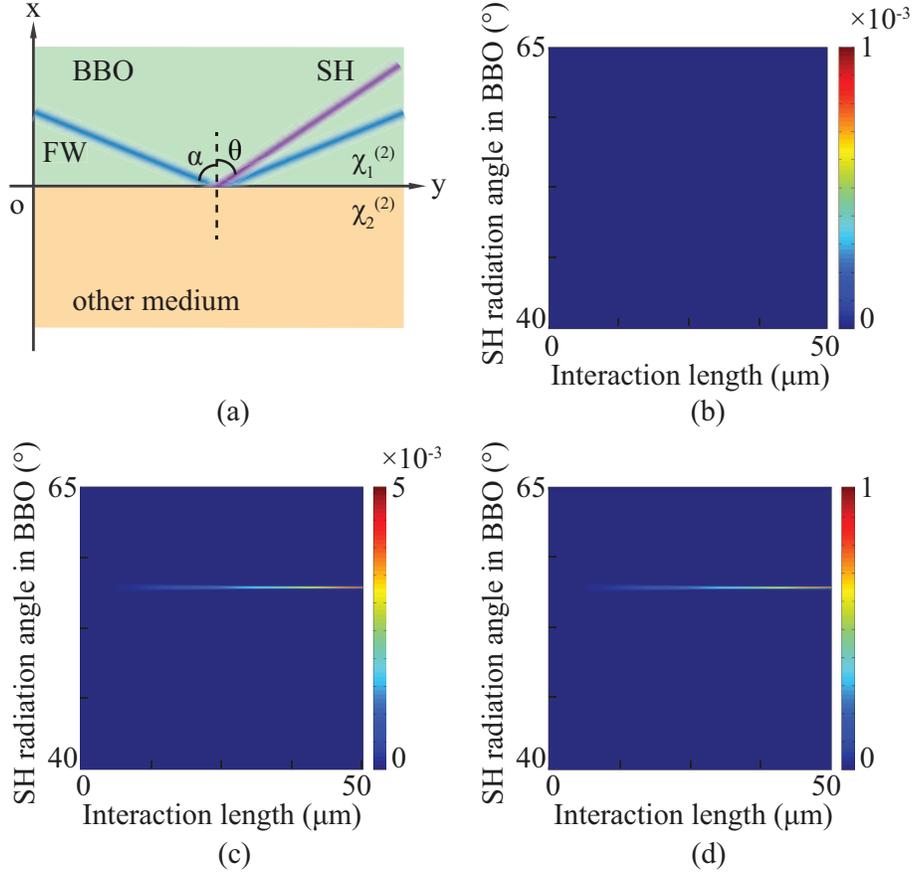}}
\caption{(a) Diagram of SH generated at the interface of BBO ($\chi^{(2)}_1$) and another medium ($\chi^{(2)}_2$). (b) Simulation result under the condition $\chi^{(2)}_1=\chi^{(2)}_2$ which presents a homogeneous nonlinear coefficient environment. (c) and (d) are intensities of SH generted on BBO boundary ($\chi^{(2)}_2=0$) and the interface of BBO and LN ($\chi^{(2)}_2=13\chi^{(2)}_1$), respectively.}
\end{figure}

The simulation results are shown in Fig. 2 with the FW wavelength of 650 nm and the incident angle $\alpha=60^\circ$. Both FW and SH are o-polarized in BBO, therefore, the nonlinear coefficient is $\chi^{(2)}_1=2d_{11}^{\mathrm{BBO}}$. The simulation result by assuming $|\chi^{(2)}_1-\chi^{(2)}_2|=0$ is presented in Fig. 2(b) as a contrast, where the intensity of SH almost equals to zero. When $\chi^{(2)}_2=0$, the distribution of SH intensity is shown in Fig. 2(c). SH generated on the BBO boundary will radiate at the particular angle $\theta=57^\circ$, which equals to that calculated by longitudinal phase-matching condition. When the medium is changed to LN, $\chi^{(2)}_2=2d_{33}^{\mathrm{LN}}\mathrm{sin}\alpha$ is about 13 times of $\chi^{(2)}_1$. As a result, the intensity of SH generated at the interface of BBO and LN is greatly enhanced as shown in Fig. 2(d). It is worth noting that the radiation angle is the same as that in Fig. 2(c) since SH is generated at the interface and always propagates in BBO side. Hence the radiation of SH will not be effected by the absorption and dispersion relation of LN. It indicates that one can employ a medium with a larger nonlinear coefficient to enhance the SH generated in UV-/VUV-transparent crystals no matter the medium itself is transparent or not.

\begin{figure}[htb]
\centerline{
\includegraphics[width=12.0cm]{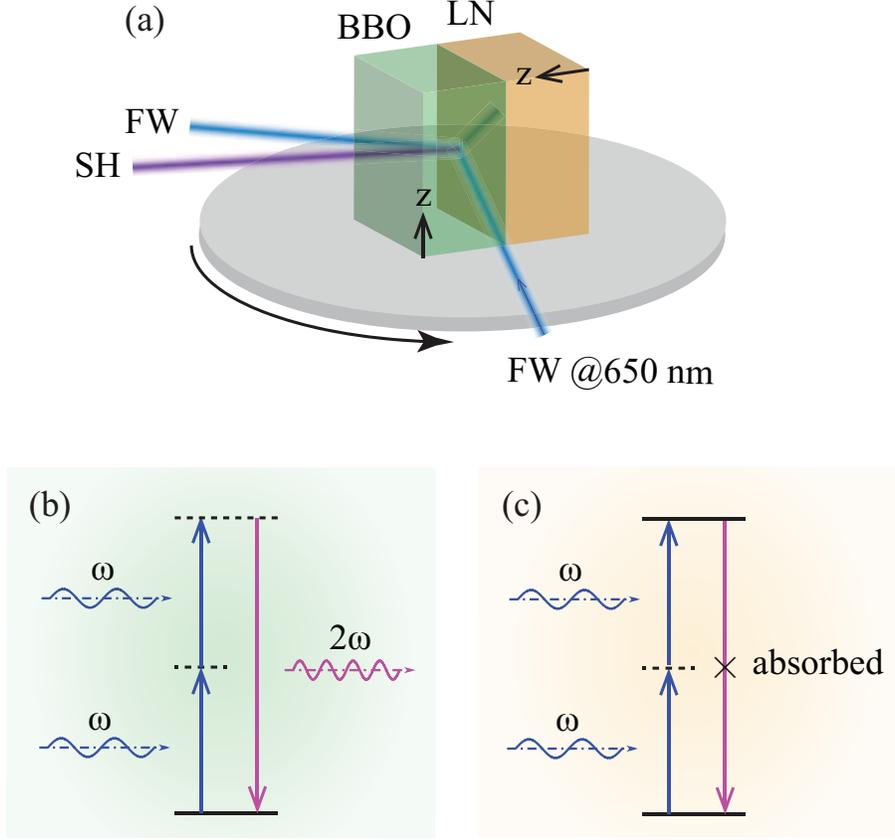}}
\caption{(a) Setup of experiment that enhanced UV second-harmonic wave at the interface of BBO and LN. (b) and (c) are energy-level discriptions of second-order nonlinear process with SH wavelength 325 nm in BBO and LN, respectively.}
\end{figure}

To demonstrate this theoretical prediction, the BBO sample and the LN sample was tightly abutted upon each other under high pressure, as shown in Fig. 3(a). The artificially fabricated interface of BBO and LN sharply modulates the nonlinear coefficient with a larger value of $|\chi^{(2)}_1-\chi^{(2)}_2|$. Similar to the above experiment, o-polarized FW with the center wavelength at 650 nm was obliquely incident onto BBO crystal and reflected at the BBO-LN interface with an angle $\alpha$.

The wavelength spectrums of SH with varying FW incident angles at the BBO-LN interface were measured by a fiber optic spectrometer, as shown in Fig. 4(a). For comparison, the SH generated on the boundary of single BBO sample was detected by the spectrometer with a same distance and angle at corresponding angles of incident FW. It's obviously that, at the same $\alpha$, the intensity of SH generated at the interface of BBO-LN was much higher than that generated at the BBO boundary. It verifies from the coupled wave equation. And in microcosmic perspective, the nonlinear polarization wave motivated by FW and propagating along the interface is the oscillation of electric dipoles. In the BBO side, the oscillating electric dipoles radiate SH at the phase-matching angle [see Fig. 3(b)]. However, in LN side, the SH at 325 nm is absorbed corresponding to inherent frequency of substance. Figure 3(c) describes this process that a atom makes a transition from its ground state to an excited state by the simultaneous absorption of two laser photons \cite{boyd2003nonlinear}, and the electric dipoles oscillation will be greatly enhanced. Although SH would not emit out of LN, the polarization on interface is intensified, which, meanwhile, enhances the SH generation in the BBO crystal.

\begin{figure}[htb]
\centerline{
\includegraphics[width=12.0cm]{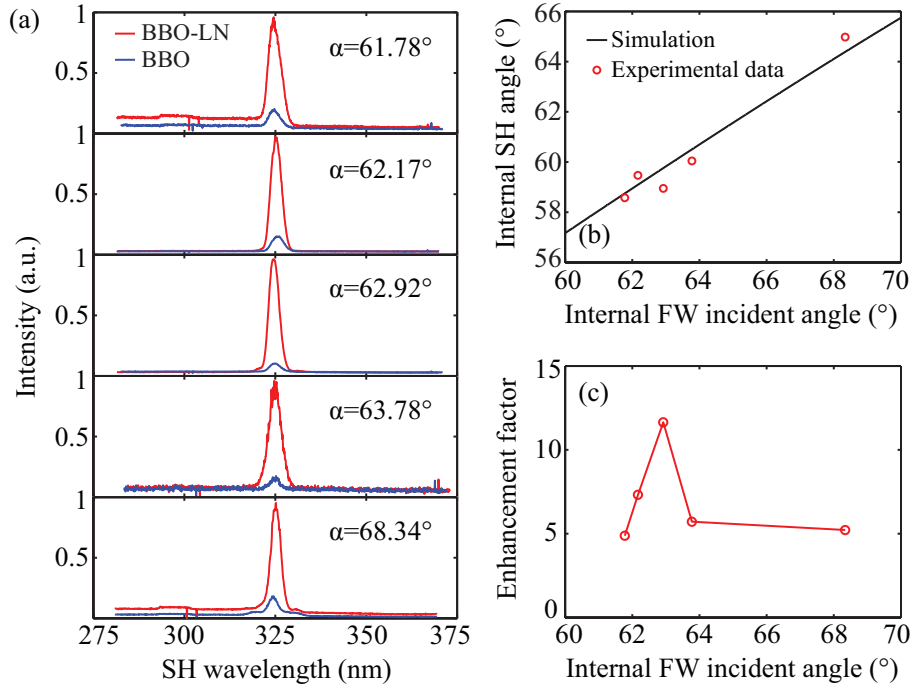}}
\caption{(a) Spectrums of SH generated on BBO boundary (blue) and interface of BBO-LN (red) with a set of FW incident angles. (b) Internal radiation angle of SH $\theta$ versus internal FW incident angle $\alpha$. Theoretical prediction (solid curve) and experimental results (symbols) are in good agreement. (c) Measured enhancement factor of SH generated at the BBO-LN interface and boundary of BBO crystal.}
\end{figure}

Figure 4(b) shows the relationship of measured SH radiation angle $\theta$ versus FW incident angle $\alpha$, which agrees well with theoretical prediction. It demonstrates the attached LN sample doesn't change the phase-matching condition. The enhancement factor of SH intensity introduced by LN was from 5 to 13 in our experiment [Fig. 4(c)] smaller than the theoretical prediction. The FW refraction from BBO to LN reduces the bump energy of this nonlinear process. Using direct bonding or crystal growth techniques, a direct interface of two crystals may be fabricated to generate enhanced SH radiation at required wavelength region.

In conclusion, the SH generated on a crystal boundary under an automatic and tunable phase matching method can be enhanced by another medium with larger nonlinear coefficient even though it is non-transparency at the SH wavelength. The intensity of SH generated on this artificially fabricated interface was about 5 to 15 times of that on single crystal boundary in our experiment. The radiation direction of SH will not be effect by the pressed crystal. It suggests potential applications in the UV even vacuum-UV spectral region.\\

This work is supported in part by the National Basic Research Program 973 of China under Grant No.2011CB808101, the National Natural Science Foundation of China under Grant Nos. 61125503, 61235009, 61205110, 61505189 and 11604318, the Innovative Foundation of Laser Fusion Research Center, and in part by Presidential Foundation of the China Academy of Engineering Physics (Grant No.201501023).

%\bibliography{UVNCR}

\begin{thebibliography}{26}%
\makeatletter
\providecommand \@ifxundefined [1]{%
 \@ifx{#1\undefined}
}%
\providecommand \@ifnum [1]{%
 \ifnum #1\expandafter \@firstoftwo
 \else \expandafter \@secondoftwo
 \fi
}%
\providecommand \@ifx [1]{%
 \ifx #1\expandafter \@firstoftwo
 \else \expandafter \@secondoftwo
 \fi
}%
\providecommand \natexlab [1]{#1}%
\providecommand \enquote  [1]{``#1''}%
\providecommand \bibnamefont  [1]{#1}%
\providecommand \bibfnamefont [1]{#1}%
\providecommand \citenamefont [1]{#1}%
\providecommand \href@noop [0]{\@secondoftwo}%
\providecommand \href [0]{\begingroup \@sanitize@url \@href}%
\providecommand \@href[1]{\@@startlink{#1}\@@href}%
\providecommand \@@href[1]{\endgroup#1\@@endlink}%
\providecommand \@sanitize@url [0]{\catcode `\\12\catcode `\$12\catcode
  `\&12\catcode `\#12\catcode `\^12\catcode `\_12\catcode `\%12\relax}%
\providecommand \@@startlink[1]{}%
\providecommand \@@endlink[0]{}%
\providecommand \url  [0]{\begingroup\@sanitize@url \@url }%
\providecommand \@url [1]{\endgroup\@href {#1}{\urlprefix }}%
\providecommand \urlprefix  [0]{URL }%
\providecommand \Eprint [0]{\href }%
\providecommand \doibase [0]{http://dx.doi.org/}%
\providecommand \selectlanguage [0]{\@gobble}%
\providecommand \bibinfo  [0]{\@secondoftwo}%
\providecommand \bibfield  [0]{\@secondoftwo}%
\providecommand \translation [1]{[#1]}%
\providecommand \BibitemOpen [0]{}%
\providecommand \bibitemStop [0]{}%
\providecommand \bibitemNoStop [0]{.\EOS\space}%
\providecommand \EOS [0]{\spacefactor3000\relax}%
\providecommand \BibitemShut  [1]{\csname bibitem#1\endcsname}%
\let\auto@bib@innerbib\@empty
%</preamble>
\bibitem [{\citenamefont {Suganuma}\ \emph {et~al.}(2002)\citenamefont
  {Suganuma}, \citenamefont {Kubo}, \citenamefont {Wakabayashi}, \citenamefont
  {Mizoguchi}, \citenamefont {Nakao}, \citenamefont {Nabekawa}, \citenamefont
  {Togashi},\ and\ \citenamefont {Watanabe}}]{suganuma2002157}%
  \BibitemOpen
  \bibfield  {author} {\bibinfo {author} {\bibfnamefont {T.}~\bibnamefont
  {Suganuma}}, \bibinfo {author} {\bibfnamefont {H.}~\bibnamefont {Kubo}},
  \bibinfo {author} {\bibfnamefont {O.}~\bibnamefont {Wakabayashi}}, \bibinfo
  {author} {\bibfnamefont {H.}~\bibnamefont {Mizoguchi}}, \bibinfo {author}
  {\bibfnamefont {K.}~\bibnamefont {Nakao}}, \bibinfo {author} {\bibfnamefont
  {Y.}~\bibnamefont {Nabekawa}}, \bibinfo {author} {\bibfnamefont
  {T.}~\bibnamefont {Togashi}}, \ and\ \bibinfo {author} {\bibfnamefont
  {S.}~\bibnamefont {Watanabe}},\ }\href@noop {} {\bibfield  {journal}
  {\bibinfo  {journal} {Optics letters}\ }\textbf {\bibinfo {volume} {27}},\
  \bibinfo {pages} {46} (\bibinfo {year} {2002})}\BibitemShut {NoStop}%
\bibitem [{\citenamefont {Togashi}\ \emph {et~al.}(2003)\citenamefont
  {Togashi}, \citenamefont {Kanai}, \citenamefont {Sekikawa}, \citenamefont
  {Watanabe}, \citenamefont {Chen}, \citenamefont {Zhang}, \citenamefont {Xu},\
  and\ \citenamefont {Wang}}]{togashi2003generation}%
  \BibitemOpen
  \bibfield  {author} {\bibinfo {author} {\bibfnamefont {T.}~\bibnamefont
  {Togashi}}, \bibinfo {author} {\bibfnamefont {T.}~\bibnamefont {Kanai}},
  \bibinfo {author} {\bibfnamefont {T.}~\bibnamefont {Sekikawa}}, \bibinfo
  {author} {\bibfnamefont {S.}~\bibnamefont {Watanabe}}, \bibinfo {author}
  {\bibfnamefont {C.}~\bibnamefont {Chen}}, \bibinfo {author} {\bibfnamefont
  {C.}~\bibnamefont {Zhang}}, \bibinfo {author} {\bibfnamefont
  {Z.}~\bibnamefont {Xu}}, \ and\ \bibinfo {author} {\bibfnamefont
  {J.}~\bibnamefont {Wang}},\ }\href@noop {} {\bibfield  {journal} {\bibinfo
  {journal} {Optics letters}\ }\textbf {\bibinfo {volume} {28}},\ \bibinfo
  {pages} {254} (\bibinfo {year} {2003})}\BibitemShut {NoStop}%
\bibitem [{\citenamefont {Kiss}\ \emph {et~al.}(2005)\citenamefont {Kiss},
  \citenamefont {Kanetaka}, \citenamefont {Yokoya}, \citenamefont {Shimojima},
  \citenamefont {Kanai}, \citenamefont {Shin}, \citenamefont {Onuki},
  \citenamefont {Togashi}, \citenamefont {Zhang}, \citenamefont {Chen} \emph
  {et~al.}}]{kiss2005photoemission}%
  \BibitemOpen
  \bibfield  {author} {\bibinfo {author} {\bibfnamefont {T.}~\bibnamefont
  {Kiss}}, \bibinfo {author} {\bibfnamefont {F.}~\bibnamefont {Kanetaka}},
  \bibinfo {author} {\bibfnamefont {T.}~\bibnamefont {Yokoya}}, \bibinfo
  {author} {\bibfnamefont {T.}~\bibnamefont {Shimojima}}, \bibinfo {author}
  {\bibfnamefont {K.}~\bibnamefont {Kanai}}, \bibinfo {author} {\bibfnamefont
  {S.}~\bibnamefont {Shin}}, \bibinfo {author} {\bibfnamefont {Y.}~\bibnamefont
  {Onuki}}, \bibinfo {author} {\bibfnamefont {T.}~\bibnamefont {Togashi}},
  \bibinfo {author} {\bibfnamefont {C.}~\bibnamefont {Zhang}}, \bibinfo
  {author} {\bibfnamefont {C.}~\bibnamefont {Chen}},  \emph {et~al.},\
  }\href@noop {} {\bibfield  {journal} {\bibinfo  {journal} {Physical review
  letters}\ }\textbf {\bibinfo {volume} {94}},\ \bibinfo {pages} {057001}
  (\bibinfo {year} {2005})}\BibitemShut {NoStop}%
\bibitem [{\citenamefont {Kiss}\ \emph {et~al.}(2008)\citenamefont {Kiss},
  \citenamefont {Shimojima}, \citenamefont {Ishizaka}, \citenamefont
  {Chainani}, \citenamefont {Togashi}, \citenamefont {Kanai}, \citenamefont
  {Wang}, \citenamefont {Chen}, \citenamefont {Watanabe},\ and\ \citenamefont
  {Shin}}]{kiss2008versatile}%
  \BibitemOpen
  \bibfield  {author} {\bibinfo {author} {\bibfnamefont {T.}~\bibnamefont
  {Kiss}}, \bibinfo {author} {\bibfnamefont {T.}~\bibnamefont {Shimojima}},
  \bibinfo {author} {\bibfnamefont {K.}~\bibnamefont {Ishizaka}}, \bibinfo
  {author} {\bibfnamefont {A.}~\bibnamefont {Chainani}}, \bibinfo {author}
  {\bibfnamefont {T.}~\bibnamefont {Togashi}}, \bibinfo {author} {\bibfnamefont
  {T.}~\bibnamefont {Kanai}}, \bibinfo {author} {\bibfnamefont
  {X.}~\bibnamefont {Wang}}, \bibinfo {author} {\bibfnamefont {C.}~\bibnamefont
  {Chen}}, \bibinfo {author} {\bibfnamefont {S.}~\bibnamefont {Watanabe}}, \
  and\ \bibinfo {author} {\bibfnamefont {S.}~\bibnamefont {Shin}},\ }\href@noop
  {} {\bibfield  {journal} {\bibinfo  {journal} {The Review of scientific
  instruments}\ }\textbf {\bibinfo {volume} {79}},\ \bibinfo {pages} {023106}
  (\bibinfo {year} {2008})}\BibitemShut {NoStop}%
\bibitem [{\citenamefont {Okazaki}\ \emph {et~al.}(2012)\citenamefont
  {Okazaki}, \citenamefont {Ota}, \citenamefont {Kotani}, \citenamefont
  {Malaeb}, \citenamefont {Ishida}, \citenamefont {Shimojima}, \citenamefont
  {Kiss}, \citenamefont {Watanabe}, \citenamefont {Chen}, \citenamefont {Kihou}
  \emph {et~al.}}]{okazaki2012octet}%
  \BibitemOpen
  \bibfield  {author} {\bibinfo {author} {\bibfnamefont {K.}~\bibnamefont
  {Okazaki}}, \bibinfo {author} {\bibfnamefont {Y.}~\bibnamefont {Ota}},
  \bibinfo {author} {\bibfnamefont {Y.}~\bibnamefont {Kotani}}, \bibinfo
  {author} {\bibfnamefont {W.}~\bibnamefont {Malaeb}}, \bibinfo {author}
  {\bibfnamefont {Y.}~\bibnamefont {Ishida}}, \bibinfo {author} {\bibfnamefont
  {T.}~\bibnamefont {Shimojima}}, \bibinfo {author} {\bibfnamefont
  {T.}~\bibnamefont {Kiss}}, \bibinfo {author} {\bibfnamefont {S.}~\bibnamefont
  {Watanabe}}, \bibinfo {author} {\bibfnamefont {C.-T.}\ \bibnamefont {Chen}},
  \bibinfo {author} {\bibfnamefont {K.}~\bibnamefont {Kihou}},  \emph
  {et~al.},\ }\href@noop {} {\bibfield  {journal} {\bibinfo  {journal}
  {Science}\ }\textbf {\bibinfo {volume} {337}},\ \bibinfo {pages} {1314}
  (\bibinfo {year} {2012})}\BibitemShut {NoStop}%
\bibitem [{\citenamefont {Pudavar}\ \emph {et~al.}(1999)\citenamefont
  {Pudavar}, \citenamefont {Joshi}, \citenamefont {Prasad},\ and\ \citenamefont
  {Reinhardt}}]{pudavar1999high}%
  \BibitemOpen
  \bibfield  {author} {\bibinfo {author} {\bibfnamefont {H.~E.}\ \bibnamefont
  {Pudavar}}, \bibinfo {author} {\bibfnamefont {M.~P.}\ \bibnamefont {Joshi}},
  \bibinfo {author} {\bibfnamefont {P.~N.}\ \bibnamefont {Prasad}}, \ and\
  \bibinfo {author} {\bibfnamefont {B.~A.}\ \bibnamefont {Reinhardt}},\
  }\href@noop {} {\bibfield  {journal} {\bibinfo  {journal} {Applied Physics
  Letters}\ }\textbf {\bibinfo {volume} {74}},\ \bibinfo {pages} {1338}
  (\bibinfo {year} {1999})}\BibitemShut {NoStop}%
\bibitem [{\citenamefont {Chen}\ \emph {et~al.}(1995)\citenamefont {Chen},
  \citenamefont {Wang}, \citenamefont {Xia}, \citenamefont {Wu}, \citenamefont
  {Tang}, \citenamefont {Wu}, \citenamefont {Wenrong}, \citenamefont {Yu},\
  and\ \citenamefont {Mei}}]{chen1995new}%
  \BibitemOpen
  \bibfield  {author} {\bibinfo {author} {\bibfnamefont {C.}~\bibnamefont
  {Chen}}, \bibinfo {author} {\bibfnamefont {Y.}~\bibnamefont {Wang}}, \bibinfo
  {author} {\bibfnamefont {Y.}~\bibnamefont {Xia}}, \bibinfo {author}
  {\bibfnamefont {B.}~\bibnamefont {Wu}}, \bibinfo {author} {\bibfnamefont
  {D.}~\bibnamefont {Tang}}, \bibinfo {author} {\bibfnamefont {K.}~\bibnamefont
  {Wu}}, \bibinfo {author} {\bibfnamefont {Z.}~\bibnamefont {Wenrong}},
  \bibinfo {author} {\bibfnamefont {L.}~\bibnamefont {Yu}}, \ and\ \bibinfo
  {author} {\bibfnamefont {L.}~\bibnamefont {Mei}},\ }\href@noop {} {\bibfield
  {journal} {\bibinfo  {journal} {Journal of applied physics}\ }\textbf
  {\bibinfo {volume} {77}},\ \bibinfo {pages} {2268} (\bibinfo {year}
  {1995})}\BibitemShut {NoStop}%
\bibitem [{\citenamefont {Wu}\ \emph {et~al.}(1996)\citenamefont {Wu},
  \citenamefont {Tang}, \citenamefont {Ye},\ and\ \citenamefont
  {Chen}}]{wu1996linear}%
  \BibitemOpen
  \bibfield  {author} {\bibinfo {author} {\bibfnamefont {B.}~\bibnamefont
  {Wu}}, \bibinfo {author} {\bibfnamefont {D.}~\bibnamefont {Tang}}, \bibinfo
  {author} {\bibfnamefont {N.}~\bibnamefont {Ye}}, \ and\ \bibinfo {author}
  {\bibfnamefont {C.}~\bibnamefont {Chen}},\ }\href@noop {} {\bibfield
  {journal} {\bibinfo  {journal} {Optical Materials}\ }\textbf {\bibinfo
  {volume} {5}},\ \bibinfo {pages} {105} (\bibinfo {year} {1996})}\BibitemShut
  {NoStop}%
\bibitem [{\citenamefont {Chen}(2004)}]{chen2004recent}%
  \BibitemOpen
  \bibfield  {author} {\bibinfo {author} {\bibfnamefont {C.}~\bibnamefont
  {Chen}},\ }\href@noop {} {\bibfield  {journal} {\bibinfo  {journal} {Optical
  Materials}\ }\textbf {\bibinfo {volume} {26}},\ \bibinfo {pages} {425}
  (\bibinfo {year} {2004})}\BibitemShut {NoStop}%
\bibitem [{\citenamefont {Kanai}\ \emph {et~al.}(2004)\citenamefont {Kanai},
  \citenamefont {Kanda}, \citenamefont {Sekikawa}, \citenamefont {Watanabe},
  \citenamefont {Togashi}, \citenamefont {Chen}, \citenamefont {Zhang},
  \citenamefont {Xu},\ and\ \citenamefont {Wang}}]{kanai2004generation}%
  \BibitemOpen
  \bibfield  {author} {\bibinfo {author} {\bibfnamefont {T.}~\bibnamefont
  {Kanai}}, \bibinfo {author} {\bibfnamefont {T.}~\bibnamefont {Kanda}},
  \bibinfo {author} {\bibfnamefont {T.}~\bibnamefont {Sekikawa}}, \bibinfo
  {author} {\bibfnamefont {S.}~\bibnamefont {Watanabe}}, \bibinfo {author}
  {\bibfnamefont {T.}~\bibnamefont {Togashi}}, \bibinfo {author} {\bibfnamefont
  {C.}~\bibnamefont {Chen}}, \bibinfo {author} {\bibfnamefont {C.}~\bibnamefont
  {Zhang}}, \bibinfo {author} {\bibfnamefont {Z.}~\bibnamefont {Xu}}, \ and\
  \bibinfo {author} {\bibfnamefont {J.}~\bibnamefont {Wang}},\ }\href@noop {}
  {\bibfield  {journal} {\bibinfo  {journal} {JOSA B}\ }\textbf {\bibinfo
  {volume} {21}},\ \bibinfo {pages} {370} (\bibinfo {year} {2004})}\BibitemShut
  {NoStop}%
\bibitem [{\citenamefont {Nakazato}\ \emph {et~al.}(2016)\citenamefont
  {Nakazato}, \citenamefont {Ito}, \citenamefont {Kobayashi}, \citenamefont
  {Wang}, \citenamefont {Chen},\ and\ \citenamefont
  {Watanabe}}]{nakazato2016phase}%
  \BibitemOpen
  \bibfield  {author} {\bibinfo {author} {\bibfnamefont {T.}~\bibnamefont
  {Nakazato}}, \bibinfo {author} {\bibfnamefont {I.}~\bibnamefont {Ito}},
  \bibinfo {author} {\bibfnamefont {Y.}~\bibnamefont {Kobayashi}}, \bibinfo
  {author} {\bibfnamefont {X.}~\bibnamefont {Wang}}, \bibinfo {author}
  {\bibfnamefont {C.}~\bibnamefont {Chen}}, \ and\ \bibinfo {author}
  {\bibfnamefont {S.}~\bibnamefont {Watanabe}},\ }\href@noop {} {\bibfield
  {journal} {\bibinfo  {journal} {Optics Express}\ }\textbf {\bibinfo {volume}
  {24}},\ \bibinfo {pages} {17149} (\bibinfo {year} {2016})}\BibitemShut
  {NoStop}%
\bibitem [{\citenamefont {Shimamura}\ \emph {et~al.}(2005)\citenamefont
  {Shimamura}, \citenamefont {V{\'\i}llora}, \citenamefont {Muramatsu},\ and\
  \citenamefont {Ichinose}}]{shimamura2005advantageous}%
  \BibitemOpen
  \bibfield  {author} {\bibinfo {author} {\bibfnamefont {K.}~\bibnamefont
  {Shimamura}}, \bibinfo {author} {\bibfnamefont {E.~G.}\ \bibnamefont
  {V{\'\i}llora}}, \bibinfo {author} {\bibfnamefont {K.}~\bibnamefont
  {Muramatsu}}, \ and\ \bibinfo {author} {\bibfnamefont {N.}~\bibnamefont
  {Ichinose}},\ }\href@noop {} {\bibfield  {journal} {\bibinfo  {journal}
  {Journal of crystal growth}\ }\textbf {\bibinfo {volume} {275}},\ \bibinfo
  {pages} {128} (\bibinfo {year} {2005})}\BibitemShut {NoStop}%
\bibitem [{\citenamefont {Kannan}\ \emph {et~al.}(2008)\citenamefont {Kannan},
  \citenamefont {Shimamura}, \citenamefont {Zeng}, \citenamefont {Kimura},
  \citenamefont {Villora},\ and\ \citenamefont
  {Kitamura}}]{kannan2008ferroelectric}%
  \BibitemOpen
  \bibfield  {author} {\bibinfo {author} {\bibfnamefont {C.}~\bibnamefont
  {Kannan}}, \bibinfo {author} {\bibfnamefont {K.}~\bibnamefont {Shimamura}},
  \bibinfo {author} {\bibfnamefont {H.}~\bibnamefont {Zeng}}, \bibinfo {author}
  {\bibfnamefont {H.}~\bibnamefont {Kimura}}, \bibinfo {author} {\bibfnamefont
  {E.}~\bibnamefont {Villora}}, \ and\ \bibinfo {author} {\bibfnamefont
  {K.}~\bibnamefont {Kitamura}},\ }\href@noop {} {\bibfield  {journal}
  {\bibinfo  {journal} {Journal of Applied Physics}\ }\textbf {\bibinfo
  {volume} {104}},\ \bibinfo {pages} {4113} (\bibinfo {year}
  {2008})}\BibitemShut {NoStop}%
\bibitem [{\citenamefont {Shimamura}\ and\ \citenamefont
  {V{\'\i}llora}(2011)}]{shimamura2011growth}%
  \BibitemOpen
  \bibfield  {author} {\bibinfo {author} {\bibfnamefont {K.}~\bibnamefont
  {Shimamura}}\ and\ \bibinfo {author} {\bibfnamefont {E.~G.}\ \bibnamefont
  {V{\'\i}llora}},\ }\href@noop {} {\bibfield  {journal} {\bibinfo  {journal}
  {Journal of Fluorine Chemistry}\ }\textbf {\bibinfo {volume} {132}},\
  \bibinfo {pages} {1040} (\bibinfo {year} {2011})}\BibitemShut {NoStop}%
\bibitem [{\citenamefont {Nomura}\ \emph {et~al.}(2011)\citenamefont {Nomura},
  \citenamefont {Ito}, \citenamefont {Ozawa}, \citenamefont {Wang},
  \citenamefont {Chen}, \citenamefont {Shin}, \citenamefont {Watanabe},\ and\
  \citenamefont {Kobayashi}}]{nomura2011coherent}%
  \BibitemOpen
  \bibfield  {author} {\bibinfo {author} {\bibfnamefont {Y.}~\bibnamefont
  {Nomura}}, \bibinfo {author} {\bibfnamefont {Y.}~\bibnamefont {Ito}},
  \bibinfo {author} {\bibfnamefont {A.}~\bibnamefont {Ozawa}}, \bibinfo
  {author} {\bibfnamefont {X.}~\bibnamefont {Wang}}, \bibinfo {author}
  {\bibfnamefont {C.}~\bibnamefont {Chen}}, \bibinfo {author} {\bibfnamefont
  {S.}~\bibnamefont {Shin}}, \bibinfo {author} {\bibfnamefont {S.}~\bibnamefont
  {Watanabe}}, \ and\ \bibinfo {author} {\bibfnamefont {Y.}~\bibnamefont
  {Kobayashi}},\ }\href@noop {} {\bibfield  {journal} {\bibinfo  {journal}
  {Optics letters}\ }\textbf {\bibinfo {volume} {36}},\ \bibinfo {pages} {1758}
  (\bibinfo {year} {2011})}\BibitemShut {NoStop}%
\bibitem [{\citenamefont {V{\'Y}llora}\ \emph {et~al.}(2009)\citenamefont
  {V{\'Y}llora}, \citenamefont {Shimamura}, \citenamefont {Sumiya},\ and\
  \citenamefont {Ishibashi}}]{vyllora2009birefringent}%
  \BibitemOpen
  \bibfield  {author} {\bibinfo {author} {\bibfnamefont {E.~G.}\ \bibnamefont
  {V{\'Y}llora}}, \bibinfo {author} {\bibfnamefont {K.}~\bibnamefont
  {Shimamura}}, \bibinfo {author} {\bibfnamefont {K.}~\bibnamefont {Sumiya}}, \
  and\ \bibinfo {author} {\bibfnamefont {H.}~\bibnamefont {Ishibashi}},\
  }\href@noop {} {\bibfield  {journal} {\bibinfo  {journal} {Optics express}\
  }\textbf {\bibinfo {volume} {17}},\ \bibinfo {pages} {12362} (\bibinfo {year}
  {2009})}\BibitemShut {NoStop}%
\bibitem [{\citenamefont {Mateos}\ \emph {et~al.}(2014)\citenamefont {Mateos},
  \citenamefont {Ram{\'\i}rez}, \citenamefont {Carrasco}, \citenamefont
  {Molina}, \citenamefont {Galisteo-L{\'o}pez}, \citenamefont {V{\'\i}llora},
  \citenamefont {de~las Heras}, \citenamefont {Shimamura}, \citenamefont
  {Lopez},\ and\ \citenamefont {Baus{\'a}}}]{mateos2014bamgf4}%
  \BibitemOpen
  \bibfield  {author} {\bibinfo {author} {\bibfnamefont {L.}~\bibnamefont
  {Mateos}}, \bibinfo {author} {\bibfnamefont {M.~O.}\ \bibnamefont
  {Ram{\'\i}rez}}, \bibinfo {author} {\bibfnamefont {I.}~\bibnamefont
  {Carrasco}}, \bibinfo {author} {\bibfnamefont {P.}~\bibnamefont {Molina}},
  \bibinfo {author} {\bibfnamefont {J.~F.}\ \bibnamefont {Galisteo-L{\'o}pez}},
  \bibinfo {author} {\bibfnamefont {E.~G.}\ \bibnamefont {V{\'\i}llora}},
  \bibinfo {author} {\bibfnamefont {C.}~\bibnamefont {de~las Heras}}, \bibinfo
  {author} {\bibfnamefont {K.}~\bibnamefont {Shimamura}}, \bibinfo {author}
  {\bibfnamefont {C.}~\bibnamefont {Lopez}}, \ and\ \bibinfo {author}
  {\bibfnamefont {L.~E.}\ \bibnamefont {Baus{\'a}}},\ }\href@noop {} {\bibfield
   {journal} {\bibinfo  {journal} {Advanced Functional Materials}\ }\textbf
  {\bibinfo {volume} {24}},\ \bibinfo {pages} {1509} (\bibinfo {year}
  {2014})}\BibitemShut {NoStop}%
\bibitem [{\citenamefont {Buchter}\ \emph {et~al.}(2001)\citenamefont
  {Buchter}, \citenamefont {Fan}, \citenamefont {Liberman}, \citenamefont
  {Zayhowski}, \citenamefont {Rothschild}, \citenamefont {Mason}, \citenamefont
  {Cassanho}, \citenamefont {Jenssen},\ and\ \citenamefont
  {Burnett}}]{Buchter:01}%
  \BibitemOpen
  \bibfield  {author} {\bibinfo {author} {\bibfnamefont {S.~C.}\ \bibnamefont
  {Buchter}}, \bibinfo {author} {\bibfnamefont {T.~Y.}\ \bibnamefont {Fan}},
  \bibinfo {author} {\bibfnamefont {V.}~\bibnamefont {Liberman}}, \bibinfo
  {author} {\bibfnamefont {J.~J.}\ \bibnamefont {Zayhowski}}, \bibinfo {author}
  {\bibfnamefont {M.}~\bibnamefont {Rothschild}}, \bibinfo {author}
  {\bibfnamefont {E.~J.}\ \bibnamefont {Mason}}, \bibinfo {author}
  {\bibfnamefont {A.}~\bibnamefont {Cassanho}}, \bibinfo {author}
  {\bibfnamefont {H.~P.}\ \bibnamefont {Jenssen}}, \ and\ \bibinfo {author}
  {\bibfnamefont {J.~H.}\ \bibnamefont {Burnett}},\ }\href {\doibase
  10.1364/OL.26.001693} {\bibfield  {journal} {\bibinfo  {journal} {Opt.
  Lett.}\ }\textbf {\bibinfo {volume} {26}},\ \bibinfo {pages} {1693} (\bibinfo
  {year} {2001})}\BibitemShut {NoStop}%
\bibitem [{\citenamefont {Ren}\ \emph {et~al.}(2013{\natexlab{a}})\citenamefont
  {Ren}, \citenamefont {Deng}, \citenamefont {Zheng}, \citenamefont {An},\ and\
  \citenamefont {Chen}}]{ren2013surface}%
  \BibitemOpen
  \bibfield  {author} {\bibinfo {author} {\bibfnamefont {H.}~\bibnamefont
  {Ren}}, \bibinfo {author} {\bibfnamefont {X.}~\bibnamefont {Deng}}, \bibinfo
  {author} {\bibfnamefont {Y.}~\bibnamefont {Zheng}}, \bibinfo {author}
  {\bibfnamefont {N.}~\bibnamefont {An}}, \ and\ \bibinfo {author}
  {\bibfnamefont {X.}~\bibnamefont {Chen}},\ }\href@noop {} {\bibfield
  {journal} {\bibinfo  {journal} {Applied Physics Letters}\ }\textbf {\bibinfo
  {volume} {103}},\ \bibinfo {pages} {021110} (\bibinfo {year}
  {2013}{\natexlab{a}})}\BibitemShut {NoStop}%
\bibitem [{\citenamefont {Li}\ \emph {et~al.}(2015)\citenamefont {Li},
  \citenamefont {Zhao}, \citenamefont {Zheng},\ and\ \citenamefont
  {Chen}}]{Li:15}%
  \BibitemOpen
  \bibfield  {author} {\bibinfo {author} {\bibfnamefont {T.}~\bibnamefont
  {Li}}, \bibinfo {author} {\bibfnamefont {X.}~\bibnamefont {Zhao}}, \bibinfo
  {author} {\bibfnamefont {Y.}~\bibnamefont {Zheng}}, \ and\ \bibinfo {author}
  {\bibfnamefont {X.}~\bibnamefont {Chen}},\ }\href {\doibase
  10.1364/OE.23.023827} {\bibfield  {journal} {\bibinfo  {journal} {Opt.
  Express}\ }\textbf {\bibinfo {volume} {23}},\ \bibinfo {pages} {23827}
  (\bibinfo {year} {2015})}\BibitemShut {NoStop}%
\bibitem [{\citenamefont {Zhang}\ \emph {et~al.}(2008)\citenamefont {Zhang},
  \citenamefont {Gao}, \citenamefont {Qi}, \citenamefont {Zhu},\ and\
  \citenamefont {Ming}}]{zhang2008nonlinear}%
  \BibitemOpen
  \bibfield  {author} {\bibinfo {author} {\bibfnamefont {Y.}~\bibnamefont
  {Zhang}}, \bibinfo {author} {\bibfnamefont {Z.}~\bibnamefont {Gao}}, \bibinfo
  {author} {\bibfnamefont {Z.}~\bibnamefont {Qi}}, \bibinfo {author}
  {\bibfnamefont {S.}~\bibnamefont {Zhu}}, \ and\ \bibinfo {author}
  {\bibfnamefont {N.}~\bibnamefont {Ming}},\ }\href@noop {} {\bibfield
  {journal} {\bibinfo  {journal} {Physical review letters}\ }\textbf {\bibinfo
  {volume} {100}},\ \bibinfo {pages} {163904} (\bibinfo {year}
  {2008})}\BibitemShut {NoStop}%
\bibitem [{\citenamefont {Ren}\ \emph {et~al.}(2012)\citenamefont {Ren},
  \citenamefont {Deng}, \citenamefont {Zheng}, \citenamefont {An},\ and\
  \citenamefont {Chen}}]{ren2012nonlinear}%
  \BibitemOpen
  \bibfield  {author} {\bibinfo {author} {\bibfnamefont {H.}~\bibnamefont
  {Ren}}, \bibinfo {author} {\bibfnamefont {X.}~\bibnamefont {Deng}}, \bibinfo
  {author} {\bibfnamefont {Y.}~\bibnamefont {Zheng}}, \bibinfo {author}
  {\bibfnamefont {N.}~\bibnamefont {An}}, \ and\ \bibinfo {author}
  {\bibfnamefont {X.}~\bibnamefont {Chen}},\ }\href@noop {} {\bibfield
  {journal} {\bibinfo  {journal} {Physical review letters}\ }\textbf {\bibinfo
  {volume} {108}},\ \bibinfo {pages} {223901} (\bibinfo {year}
  {2012})}\BibitemShut {NoStop}%
\bibitem [{\citenamefont {Ren}\ \emph {et~al.}(2013{\natexlab{b}})\citenamefont
  {Ren}, \citenamefont {Deng}, \citenamefont {Zheng}, \citenamefont {An},\ and\
  \citenamefont {Chen}}]{ren2013enhanced}%
  \BibitemOpen
  \bibfield  {author} {\bibinfo {author} {\bibfnamefont {H.}~\bibnamefont
  {Ren}}, \bibinfo {author} {\bibfnamefont {X.}~\bibnamefont {Deng}}, \bibinfo
  {author} {\bibfnamefont {Y.}~\bibnamefont {Zheng}}, \bibinfo {author}
  {\bibfnamefont {N.}~\bibnamefont {An}}, \ and\ \bibinfo {author}
  {\bibfnamefont {X.}~\bibnamefont {Chen}},\ }\href@noop {} {\bibfield
  {journal} {\bibinfo  {journal} {Optics letters}\ }\textbf {\bibinfo {volume}
  {38}},\ \bibinfo {pages} {1993} (\bibinfo {year}
  {2013}{\natexlab{b}})}\BibitemShut {NoStop}%
\bibitem [{\citenamefont {Sheng}\ \emph {et~al.}(2012)\citenamefont {Sheng},
  \citenamefont {Roppo}, \citenamefont {Kalinowski},\ and\ \citenamefont
  {Krolikowski}}]{sheng2012role}%
  \BibitemOpen
  \bibfield  {author} {\bibinfo {author} {\bibfnamefont {Y.}~\bibnamefont
  {Sheng}}, \bibinfo {author} {\bibfnamefont {V.}~\bibnamefont {Roppo}},
  \bibinfo {author} {\bibfnamefont {K.}~\bibnamefont {Kalinowski}}, \ and\
  \bibinfo {author} {\bibfnamefont {W.}~\bibnamefont {Krolikowski}},\
  }\href@noop {} {\bibfield  {journal} {\bibinfo  {journal} {Optics letters}\
  }\textbf {\bibinfo {volume} {37}},\ \bibinfo {pages} {3864} (\bibinfo {year}
  {2012})}\BibitemShut {NoStop}%
\bibitem [{\citenamefont {Zhao}\ \emph {et~al.}(2016)\citenamefont {Zhao},
  \citenamefont {Zheng}, \citenamefont {Ren}, \citenamefont {An}, \citenamefont
  {Deng},\ and\ \citenamefont {Chen}}]{Zhao:16}%
  \BibitemOpen
  \bibfield  {author} {\bibinfo {author} {\bibfnamefont {X.}~\bibnamefont
  {Zhao}}, \bibinfo {author} {\bibfnamefont {Y.}~\bibnamefont {Zheng}},
  \bibinfo {author} {\bibfnamefont {H.}~\bibnamefont {Ren}}, \bibinfo {author}
  {\bibfnamefont {N.}~\bibnamefont {An}}, \bibinfo {author} {\bibfnamefont
  {X.}~\bibnamefont {Deng}}, \ and\ \bibinfo {author} {\bibfnamefont
  {X.}~\bibnamefont {Chen}},\ }\href {\doibase 10.1364/OE.24.012825} {\bibfield
   {journal} {\bibinfo  {journal} {Opt. Express}\ }\textbf {\bibinfo {volume}
  {24}},\ \bibinfo {pages} {12825} (\bibinfo {year} {2016})}\BibitemShut
  {NoStop}%
\bibitem [{\citenamefont {Boyd}(2003)}]{boyd2003nonlinear}%
  \BibitemOpen
  \bibfield  {author} {\bibinfo {author} {\bibfnamefont {R.~W.}\ \bibnamefont
  {Boyd}},\ }\href@noop {} {\emph {\bibinfo {title} {Nonlinear optics}}}\
  (\bibinfo  {publisher} {Academic press},\ \bibinfo {year} {2003})\BibitemShut
  {NoStop}%
\end{thebibliography}
%

\end{document}